\documentclass[twocolumn]{aastex62}

\hypersetup{breaklinks}   

\usepackage{amsmath}

\received{\today}
\submitjournal{ApJ}

\shorttitle{Fast solar wind origin}
\shortauthors{W\'ojcik et al.}
\newcommand{\beqa}{\begin{eqnarray}}
\newcommand{\eeqa}{\end{eqnarray}}

\begin{document}

\title{\bf Two-fluid numerical simulations of the origin of the fast solar wind} 

\correspondingauthor{D. W\'ojcik}
\email{dwojcik@kft.umcs.lublin.pl}

\author{D. W\'ojcik}
\affil{Group of Astrophysics, Institute of Physics, University of M. Curie-Sk{\l}odowska, \\
ul. Radziszewskiego 10, 20-031 Lublin, Poland}

\author{B. Ku\'zma}
\affiliation{Group of Astrophysics, Institute of Physics, University of M. Curie-Sk{\l}odowska, \\
ul. Radziszewskiego 10, 20-031 Lublin, Poland}

\author{K. Murawski}
\affiliation{Group of Astrophysics, Institute of Physics, University of M. Curie-Sk{\l}odowska, \\
ul. Radziszewskiego 10, 20-031 Lublin, Poland}

\author{A.K. Srivastava}
\affiliation{Department of Physics, Indian Institute of Technology (Banaras Hindu University), \\
Varanasi-221005, India}

\begin{abstract}
With the use of our JOANNA code, which solves radiative equations for 
ion + electron and neutral fluids, we perform realistic 2.5D numerical simulations of 
plasma outflows 
associated with the solar granulation. 
These outflows exhibit physical quantities 
consistent to the order of magnitude with the observational 
findings 
for mass and energy losses in the upper chromosphere, transition region and inner corona, and they may originate the fast solar wind.
\end{abstract}

\keywords{Sun: activity - Sun: corona - Sun: solar wind - magnetohydrodynamics (MHD) - methods: numerical}

\section{Introduction} \label{sec:intro}
The solar wind is a stream of energized, charged particles, primarily electrons and protons from hydrogen, along with atomic nuclei like helium, alpha particles, flowing outward from the Sun \citep{Parker1965,Biermann1951}. 
There is a fast, widely uniform wind, emanating from polar coronal holes and at the distance of 1 AU from the Sun traveling at about 750 km s$^{-1}$, and a slow, sporadic one, pouring from active equatorial regions and moving with about half smaller speed. 
The magnetic field lines stretch out radially in coronal holes, and do not loop directly back to the Sun, providing an open path for the fast plasma to escape the gravity grasp. 
As the corona expands, these winds must be replaced by plasma moving up from below to feed them. 

In the modern era of high resolution space-born and ground-based observations, 
special attention has been paid for studying the origin of plasma outflows which form the solar wind. 
The early models of the wind assumed the inner corona as its origin \citep[e.g.][]{Tu1987}.
Recently, \cite{Tu2005} detected such outflows in coronal funnels at altitudes between 5 Mm and 20 Mm above the photosphere, and found that they reach the speed of up to 10 km s$^{-1}$ at the height of 20 Mm. 
The emphasis was also given on searching for the outflows in the chromosphere/transition region \citep[e.g.][]{Marsch2008,
McIntosh2012,Yang2013,Kayshap2015}. 
It was found that these outflows can be generated by variety of jets \citep{
WedemeyerBhm2012,Kayshap2013,Tian2014,MartnezSykora2017}, and injection of often twisted magnetic field and its subsequent reconnection may also contribute to the bulk plasma outflows in coronal holes
\citep[e.g.][]{Krieger1973,Zirker1977,Yang2013,Kayshap2013}.

It was showed theoretically that magnetohydrodynamic waves may be responsible for providing the momentum to the upwardly moving plasma \citep[e.g.][]{Ofman2005,Suzuki2005,Srivastava2006,Marsch2006,
Arber2016}. 
Among others, 
\cite{Hollweg1986}, \cite{Kudoh1999}, \cite{Matsumoto2012} concluded that Alfv\'en waves possess a potential to drive plasma outflows. 
\cite{He2008} developed a model of plasma outflows in the coronal funnels, which includes Alfv\'en waves \citep{Ofman1995}. 
Along this line of investigation, \cite{Yang2016} found that Alfv\'en waves are able to form fast plasma outflows. 
See also \cite{Ofman1995} and \cite{Shestov2017} for a similar analysis. 
However, Alfv\'en waves are difficult for detection, particularly when high-frequency waves are concerned \citep{Srivastava2017}; in an inhomogeneous and structured medium, these waves can experience reflection, mode coupling, phase-mixing and resonant absorption \citep{Ofman1995, Nakariakov1997, Zaqarashvili2006, Goossens2012
}. 

Despite the above mentioned achievements, the origin of the solar wind still remains one of the central issues of heliophysics.
We investigate here the role of granulation in generation of chromospheric ejecta and associated plasma outflows in coronal holes. 
We are motivated by the fact that a base of the corona is filled with dynamic jets propelled from below the transition region upwards at speeds of about 25 km s$^{-1}$ into higher layers, and carry a significant amount of momentum \citep[e.g.][]{Sterling2000,Zaqarashvili2009}. 

This paper is organized as follows. A physical model is presented in Sect. 2 and the corresponding numerical results are shown in Sect. 3. 
Our paper is concluded by discussion and summary of the numerical results in Sect. 4.
\section{Two-fluid model of a partially-ionized coronal hole}\label{sec:atm_model}
We consider a solar coronal hole that is magnetically structured and gravitationally stratified, 
and its dynamics is described by 2-fluid equations for ions + electrons treated as one fluid and neutrals regarded as second fluid. These equations can be written as follows:  

\begin{equation}
\frac{\partial \varrho_{\rm n}}{\partial t}+\nabla\cdot(\varrho_{\rm n} \mathbf{V}_{\rm n}) = 0\,,
\label{eq:neutral_continuity}
\end{equation}
\begin{equation}
\frac{\partial \varrho_{\rm i}}{\partial t}+\nabla\cdot(\varrho_{\rm i} \mathbf{V}_{\rm i}) = 0\,,
\label{eq:ion_continuity}
\end{equation}
\begin{equation}
\begin{split}
\frac{\partial (\varrho_{\rm n} \mathbf{V}_{\rm n})}{\partial t}+
\nabla \cdot (\varrho_{\rm n} \mathbf{V}_{\rm n} \mathbf{V}_{\rm n}+p_{\rm n} \mathbf{I}) = \\
\alpha_c({\bf V}_{\rm i}-{\bf V}_{\rm n}) + \varrho_{\rm n} \mathbf{g}\,,
\end{split}
\label{eq:neutral_momentum}
\end{equation}
\begin{equation}
\begin{split}
\frac{\partial  (\varrho_{\rm i} \mathbf{V}_{\rm i})}{\partial t}+
\nabla \cdot (\varrho_{\rm i} \mathbf{V}_{\rm i} \mathbf{V}_{\rm i}+p_{\rm ie} \mathbf{I}) = \\
\frac{1}{\mu}(\nabla \times \mathbf{B}) \times \mathbf{B} +  \alpha_c({\bf V}_{\rm n}-{\bf V}_{\rm i}) +\varrho_{\rm i} \mathbf{g}\,,
\end{split}
\label{eq:ion_momentum}
\end{equation}
\begin{equation}
\frac{\partial \mathbf{B}}{\partial t} = \nabla \times (\mathbf{V_{\rm i} \times }\mathbf{B})\,, \hspace{3mm} \nabla \cdot{\mathbf B}=0\,,
\label{eq:ions_induction}\\
\end{equation}
\begin{equation}
\begin{split}
\frac{\partial E_{\rm n}}{\partial t}+\nabla\cdot[(E_{\rm n}+p_{\rm n})\mathbf{V}_{\rm n}] = \alpha_c{\bf V}_{\rm n}({\bf V}_{\rm i}-{\bf V}_{\rm n})\\
+Q_{\rm n} ^{\rm in} + q_{\rm n} + \varrho_{\rm n} \mathbf{g} \cdot \mathbf{V}_{\rm n}\,,
\end{split}
\label{eq:neutral_energy}
\end{equation}
\begin{equation}
\begin{split}
\frac{\partial E_{\rm i}}{\partial t}+\nabla\cdot\left[\left(E_{\rm i}+p_{\rm ie} + \frac{{\bf B}^2}{2\mu} \right)\mathbf{V}_{\rm i}-\mathbf{B}(\mathbf{V}\cdot \mathbf{B})\right] = \\
\alpha_c {\bf V}_{\rm i}({\bf V}_{\rm n}-{\bf V}_{\rm i}) +Q_{\rm i}^{\rm in} + Q^i_{R} + q_{\rm i} + \varrho_{\rm i} \mathbf{g}  \cdot \mathbf{V}_{\rm i}\,, 
\end{split}
\label{eq:ion_energy}
\end{equation}

%
where the heat production terms are 
%
\begin{equation}
Q_{\rm n}^{\rm in} = \alpha_c ( \Delta \tilde{V}+ \Delta \tilde{T}  ) \,,
\end{equation}
\begin{equation}
Q_{\rm i}^{\rm in} = \alpha_c ( \Delta \tilde{V} - \Delta \tilde{T} )
\end{equation}
{\bf with}
\begin{equation}
\Delta \tilde{V} = \frac{1}{2} |{\mathbf V}_{\rm i}-{\mathbf V}_{\rm n} |^2 \,,
\end{equation}With the use of our JOANNA code, which solvesradiativeequations for ion + electron and neutralfluids,  we perform realistic 2.5D numerical simulations of plasma outflows associated with the solargranulation.  These outflows exhibit physical quantities consistentto the order of magnitude with theobservational findings  for  mass and  energy losses in  the  upper  chromosphere, transition  region andinner corona, and they may originate the fast solar wind.
\begin{equation}
\Delta \tilde{T}=\frac{3k_{\rm B}}{m_{\rm H}(\mu_{\rm i}+\mu_{\rm n})}\left(T_{\rm i}-T_{\rm n} \right )\,,
\end{equation}
and the energy densities are given by
\begin{equation}
E_{\rm n} = \frac{p_{\rm n}}{\gamma-1} + \frac{\varrho_{\rm n}\mathbf{V}_{\rm n}^2}{2}\,,
\end{equation}
\begin{equation}
E_{\rm i} = \frac{p_{\rm ie}}{\gamma-1} + \frac{\varrho_{\rm i}\mathbf{V}_{\rm i}^2}{2} + 
\frac{\mathbf{B}^2}{2\mu}\, .
\end{equation}
%
Here subscripts $_{\rm i}$, $_{\rm n}$ and $_{\rm e}$ correspond to ions, neutrals and electrons, respectively. 
The symbols $\varrho_{\rm i,n}$ denote mass densities, ${\bf V}_{\rm i,n}$ velocities, 
$p_{\rm ie,n}$ ion+electron and neutral gas pressures, 
${\bf B}$ is magnetic field and $T_{\rm i,n}$ are temperatures specified by ideal gas laws, 
\begin{equation}
p_{\rm n}=\frac{k_{\rm B}}{m_{\rm H}\mu_{\rm n}}\varrho_{\rm n}T_{\rm n}\,, \hspace*{1.0cm} p_{\rm ie}=\frac{2k_{\rm B}}{m_{\rm H}\mu_{\rm i}}\varrho_{\rm i}T_{\rm i}\,.
\label{eq:pressures}
\end{equation}
A gravity vector is ${\bf g} = [0, -g, 0]$ with its magnitude $g = 274.78$ m s$^{-2}$, 
$\alpha_{\rm c}$ is the coefficient of collisions between ion and neutral particles 
\citep[e.g.][and references cited therein]{Oliver2016,Ballester2018}, 
$Q^{\rm i}_{R}$ 
radiative losses term which is implemented here in the framework of \cite{2012SoPh..277....3A} in the low atmospheric regions and of thin radiation \citep{1972SoPh...23...78M} in the top atmospheric layers, 
$q_{\rm i,n}$ are thermal conduction terms \citep{1962pfig.book.....S}, 
$\mu_{\rm i}=0.29$ and $\mu_{\rm n}=1.21$ are the mean masses of respectively ions and neutrals, 
which are taken from the OPAL solar abundance model \citep[e.g.][]{Vogler2004}, 
$m_{\rm H}$ is the hydrogen mass, $k_{\rm B}$ is the Boltzmann constant, 
$\gamma=1.4$ is the specific heats ratio, and $\mu$ is magnetic permeability of the medium. 
The other symbols have their standard meaning. \\
%
%
%
We consider the case of $z-$invariant system and start our simulations at $t=0$ s 
with the hydrostatic equilibrium being supplemented by transversal and vertical magnetic field given as 
$\mathbf{B}=[B_{x},B_{y},B_{z}]=[0,B_{0},B_{0}]$, where $B_{0}=5/\sqrt{2}$ Gs. 
The transversal component, $B_z$ results in Alfv\'en waves being linearly coupled to magnetoacoustic waves. 
The presence of Alfv\'en waves is essential in the model, as in the nonlinear regime they are capable of 
driving vertical flow 
\citep[e.g.][]{
Hollweg1986,Murawski1992,Shestov2017}. 

In the framework of the implemented magnetic field model we set at $t=0$ s identical hydrostatic temperature 
for ions and neutrals, $T_{\rm i} (y) = T_{\rm n} (y) = T (y)$ 
\citep{Oliver2016,MartnezGmez2016,MartnezGmez2017,Soler2017,Srivastava2018}. 
This temperature is determined by the semi-empirical model of \cite{Avrett2008} that is extrapolated into the corona. 

%
%

The hydrostatic equilibrium is restructured in time by the solar granulation. 
This granulation appears naturally in the convection zone which is convectively unstable. 
First signs of granulation are seen already after about $5$ min from the start of the simulations 
with a fully developed state occurring after about 3000 s of the solar time. 
\section{Numerical simulations of 2-fluid plasma outflows}\label{sec:num_sim_MHD}

\begin{figure}
\vspace{1cm}
\begin{center}
\includegraphics[width=\columnwidth]{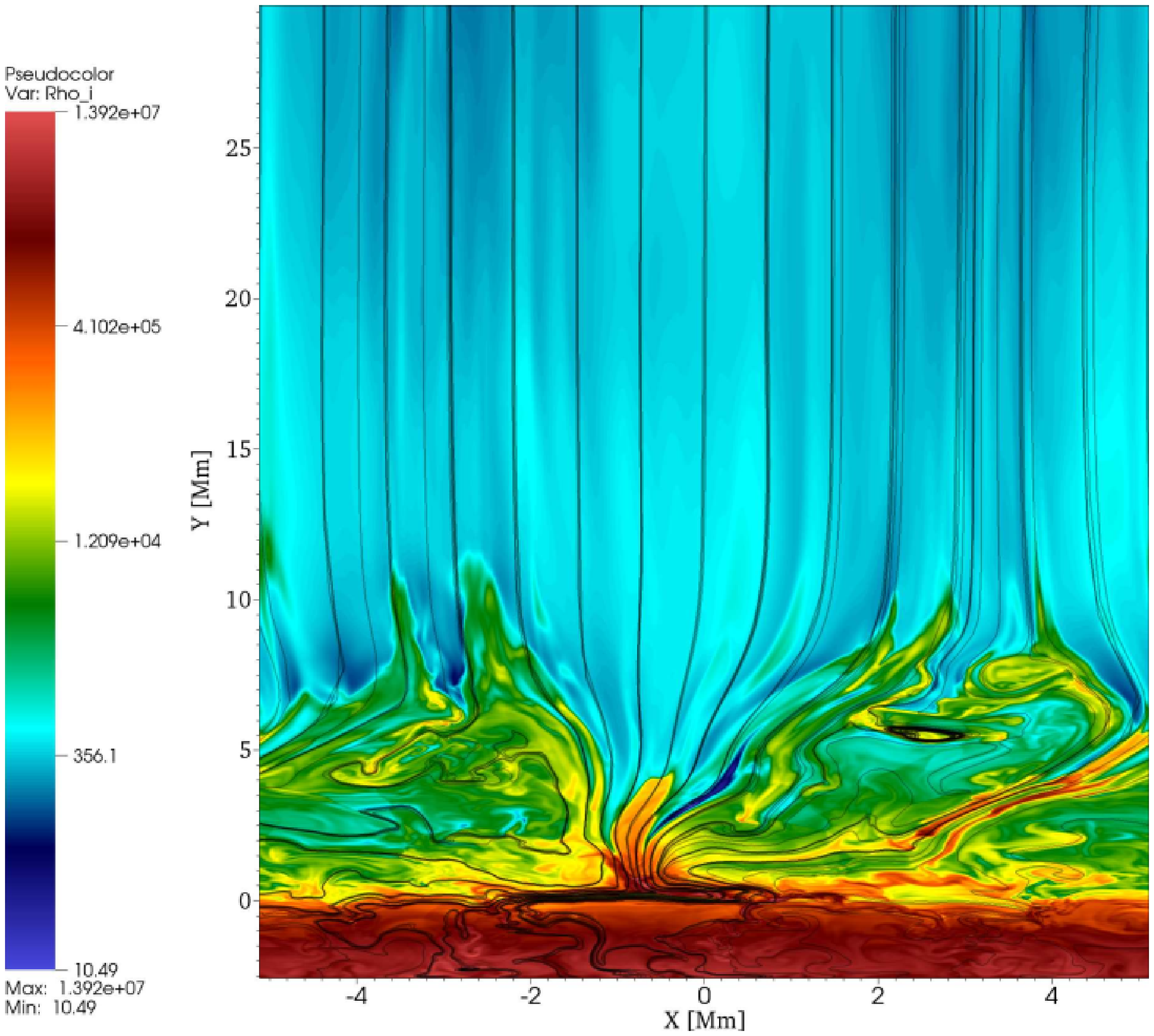}\\
\vspace*{1.0cm}\includegraphics[width=\columnwidth]{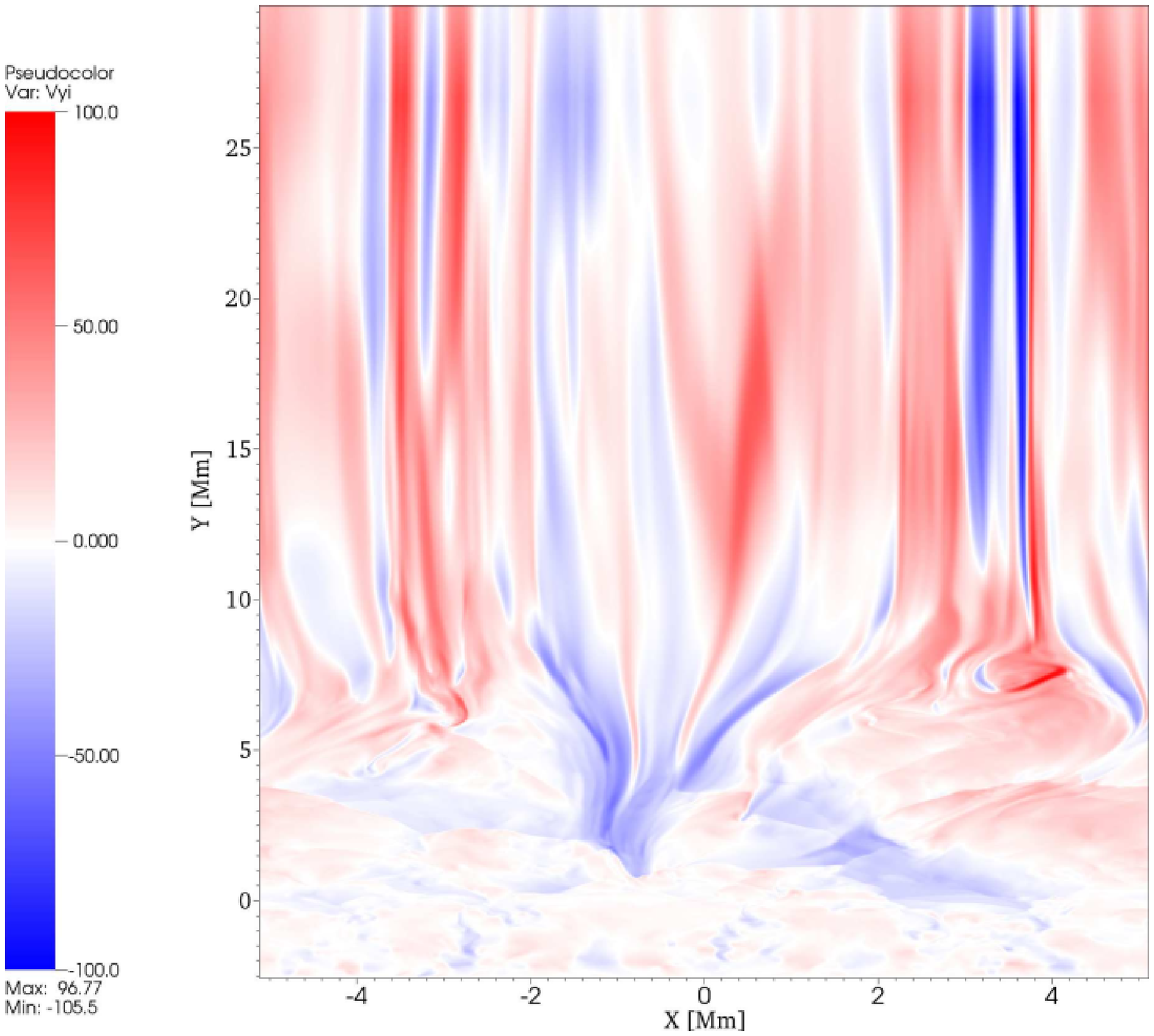}\\
\caption{
Typical spatial profiles of  $\log \varrho_{\rm i}$ overlayed by magnetic field lines (solid lines) (top) and vertical component of velocity $V_{{\rm i}\, y}(x,y)$ (bottom). Ion mass density is given in units of $10^{-18}$ g cm$^{-3}$ and $V_{{\rm i}\, y}$ is expressed in units of 1 km~s$^{-1}$. 
} 
\label{fig:temperature}
\end{center}
\end{figure}

To solve 2-fluid equations numerically, we use JOANNA code \citep{
2018MNRAS.481..262W}. 
We set in our numerical experiments the Courant-Friedrichs-Lewy number equal to $0.9$ 
and choose a second-order accuracy in space and a four stage, third-order strong stability preserving Runge-Kutta method \citep{Durran2010} for integration in time, supplemented by adopting 
the Harten-Lax-van Leer Discontinuities (HLLD) approximate Riemann solver \citep{Miyoshi2005} and Global Lagrange Multiplier (GLM) method of \cite{Dedner2002}. 
The simulation box extends from the convection zone (2.56 Mm 
below the bottom of the photosphere) to the corona (up to 30 Mm above the photosphere) in $y-$direction and horizontally from $x=-2.56$ Mm to $x=2.56$ Mm. 
This box is divided into several patches. 
The bottom region, 
specified by $-2.56$ Mm $\leq y \leq 7.68$ Mm, 
is covered by the $1024 \times 1024$ identical cells, 
leading to the spatial resolution of $10$ km. 
Within the 
layer of $7.68$ Mm $\leq y \leq 30$ Mm, 
we implement several patches of progressively larger cells along $y-$direction. 
At the left- and right- sides of the simulation box we set periodic boundary conditions, 
while the top and bottom ghost cells are filled by plasma quantities equal to their equilibrium values. 
The layer for the optical depth greater than 10
is additionally heated by implementing extra source term in the energy equation of ions that balances the 
energy
losses there.
%

%
%
\begin{figure}
\begin{center}
\includegraphics[width=\columnwidth]{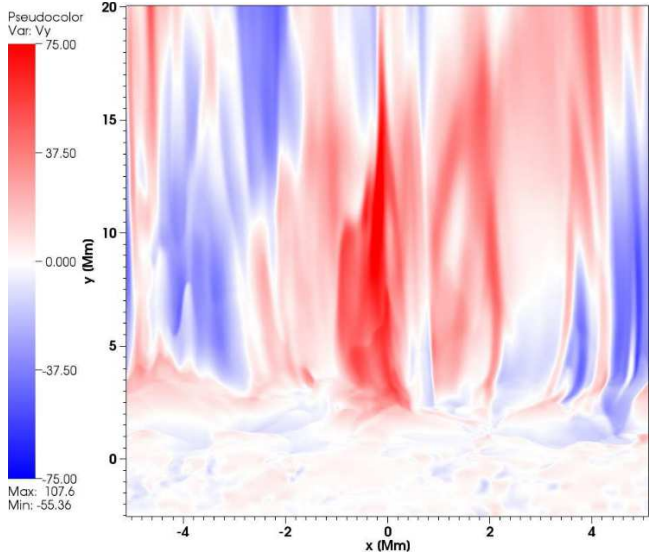}
\caption{
Spatial profile of vertical component $V_{{\rm i}\, y}(x,y)$, expressed in units of 1 km~s$^{-1}$ for the spatial resolution of $20$ km x $20$ km. 
}
\label{fig:Vyi_20x20}
\end{center}
\end{figure}
\begin{figure}
\begin{center}
\includegraphics[width=\columnwidth]{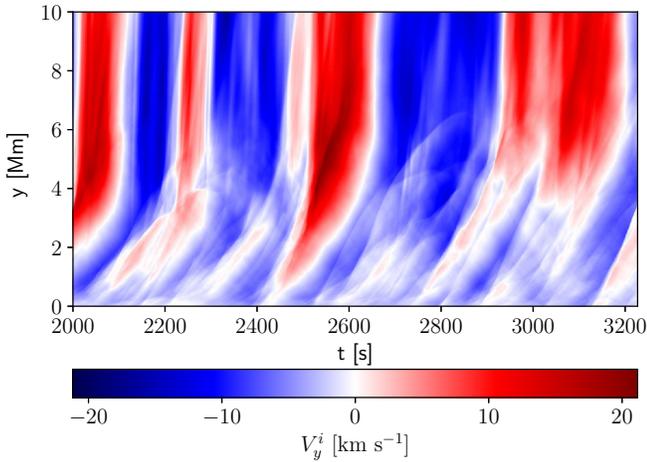}
\caption{
Time-distance plot of horizontally averaged vertical component of ion velocity, $<V_{{\rm i}\, y}(y,t)>$. 
}
\label{fig:td-plot}
\end{center}
\end{figure}

Figure~1 (colormap) shows typical spatial profiles of $\log \varrho_{\rm i}$ (top) and vertical component of ion velocity (bottom).  
The granulation excites a wide range of waves deep in the photosphere. Some of these waves steepen into shocks while propagating upwards. This steepening results from wave amplitude growth with height, and chromospheric jets are excited (top). 
The plasma above the apices of these jets 
moves upwards reaching its maximum speed of around
$100$~km s$^{-1}$ (bottom). 
The outflowing plasma essentially follows open magnetic field lines (black lines) of magnetic funnels that are formed by the granulation. The footpoints of these funnels are rooted deep in the photosphere between granules, while higher up, above the transition region, the magnetic field lines remain essentially vertical. 
The plasma outflows form strand-like structures along magnetic field lines in the corona, with subsiding plasma taking place at lower altitudes. 

We have run the code with extra non-adiabatic terms such as thermal conduction and magnetic
diffusion included along with radiation. However, due to the computational effort we have not yet
obtained satisfactory results. 
We have also run the code with the spatial resolution of 20 km x 20 km and the results are showed in Fig. \ref{fig:Vyi_20x20}. It follows from them that the results are close to these shown in Fig. \ref{fig:temperature} (bottom), which confirms that the chosen spatial resolution of 10 km x 10 km is sufficient to resolve plasma outflows. 


%
\begin{figure}
\begin{center}
\includegraphics[width=\columnwidth]{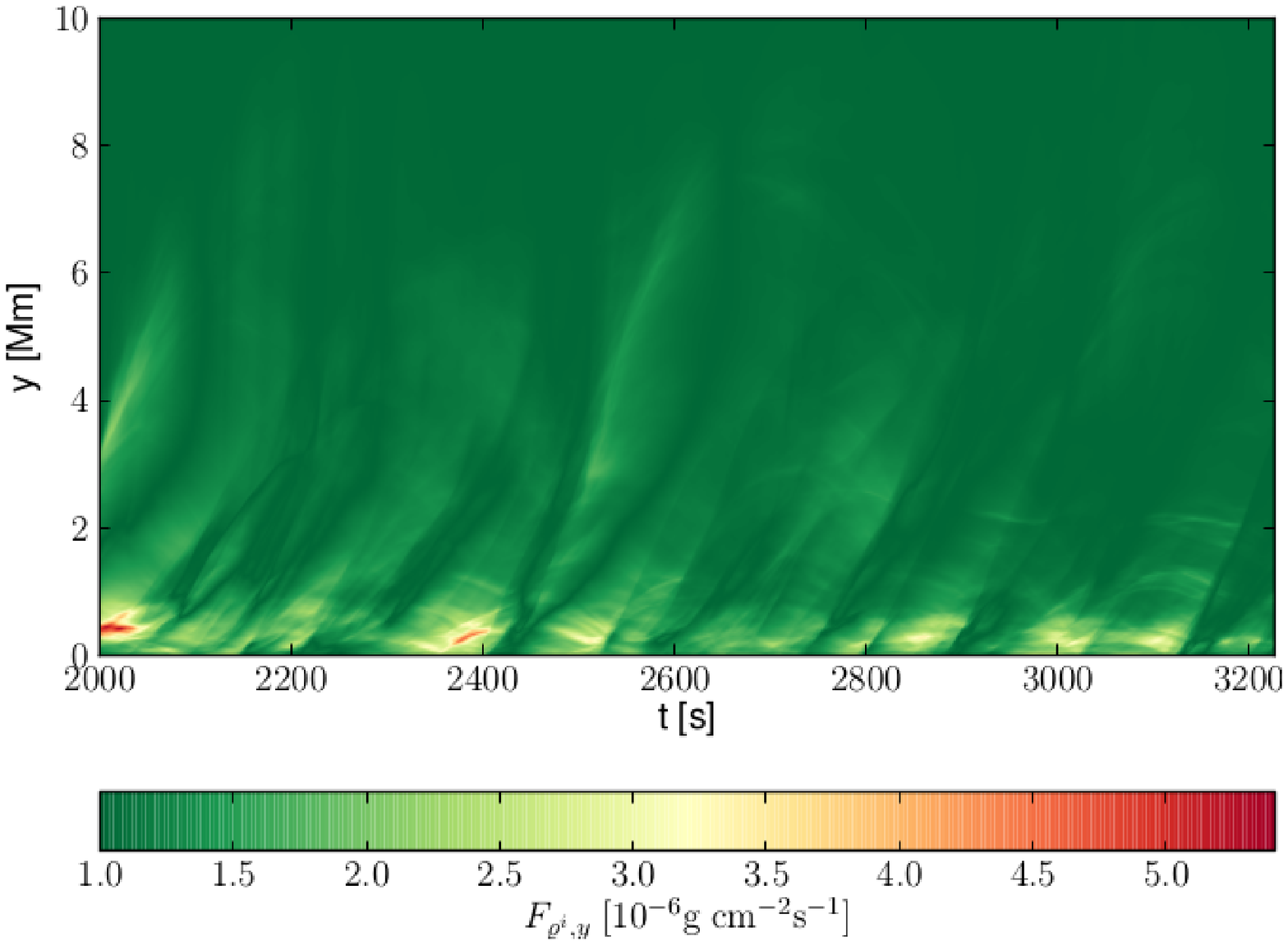}\\
\vspace*{1.0cm}
\includegraphics[width=\columnwidth]{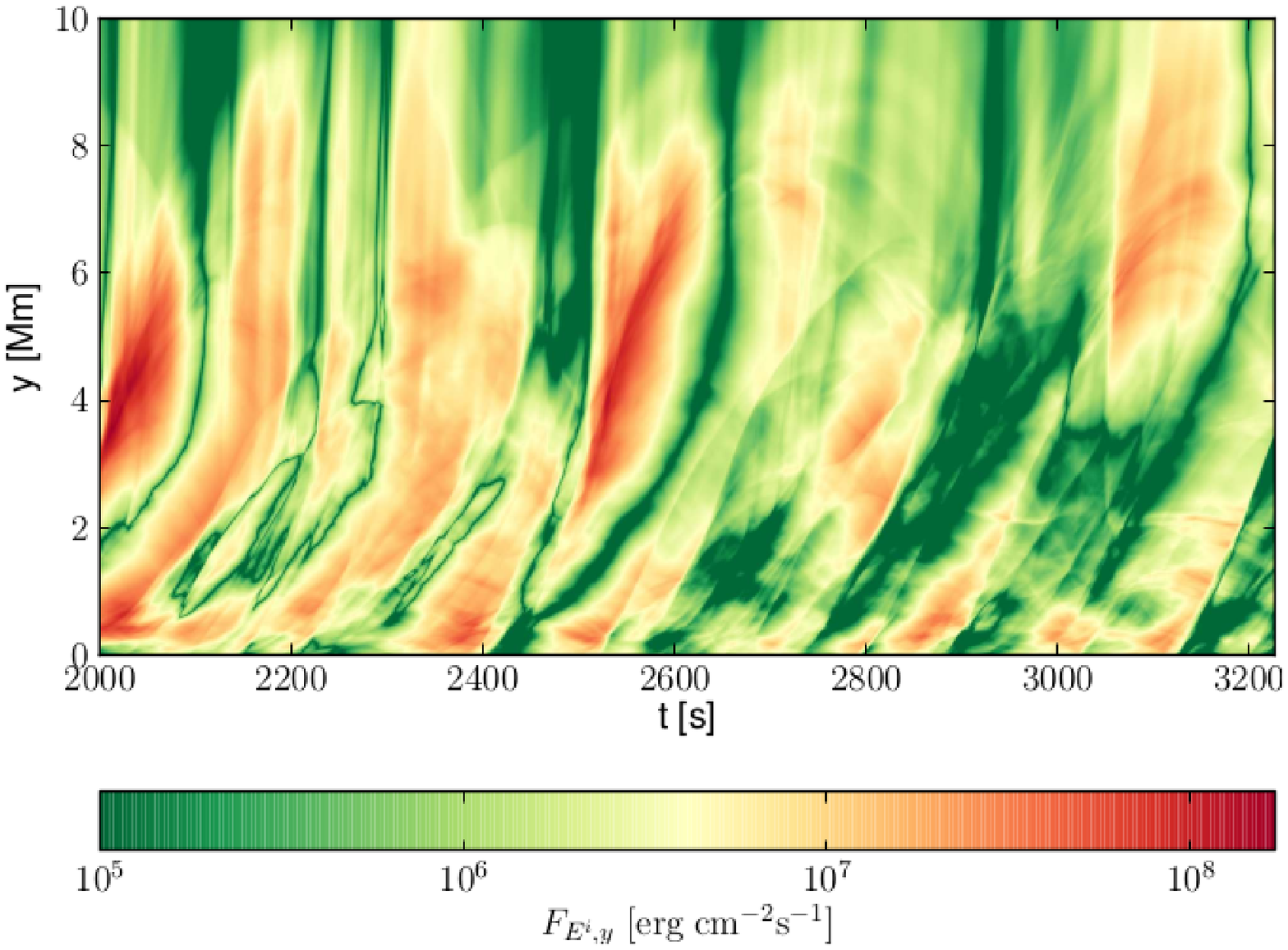}\\
\caption{
Time-distance plots of the horizontally averaged total vertical mass, $<F_{m}(y,t)>$, (top) 
and energy, $<F_{E}(y,t)>$, (bottom) fluxes. 
}
\label{fig:energyflux}
\end{center}
\end{figure}
\begin{figure}
\begin{center}
\includegraphics[width=\columnwidth]{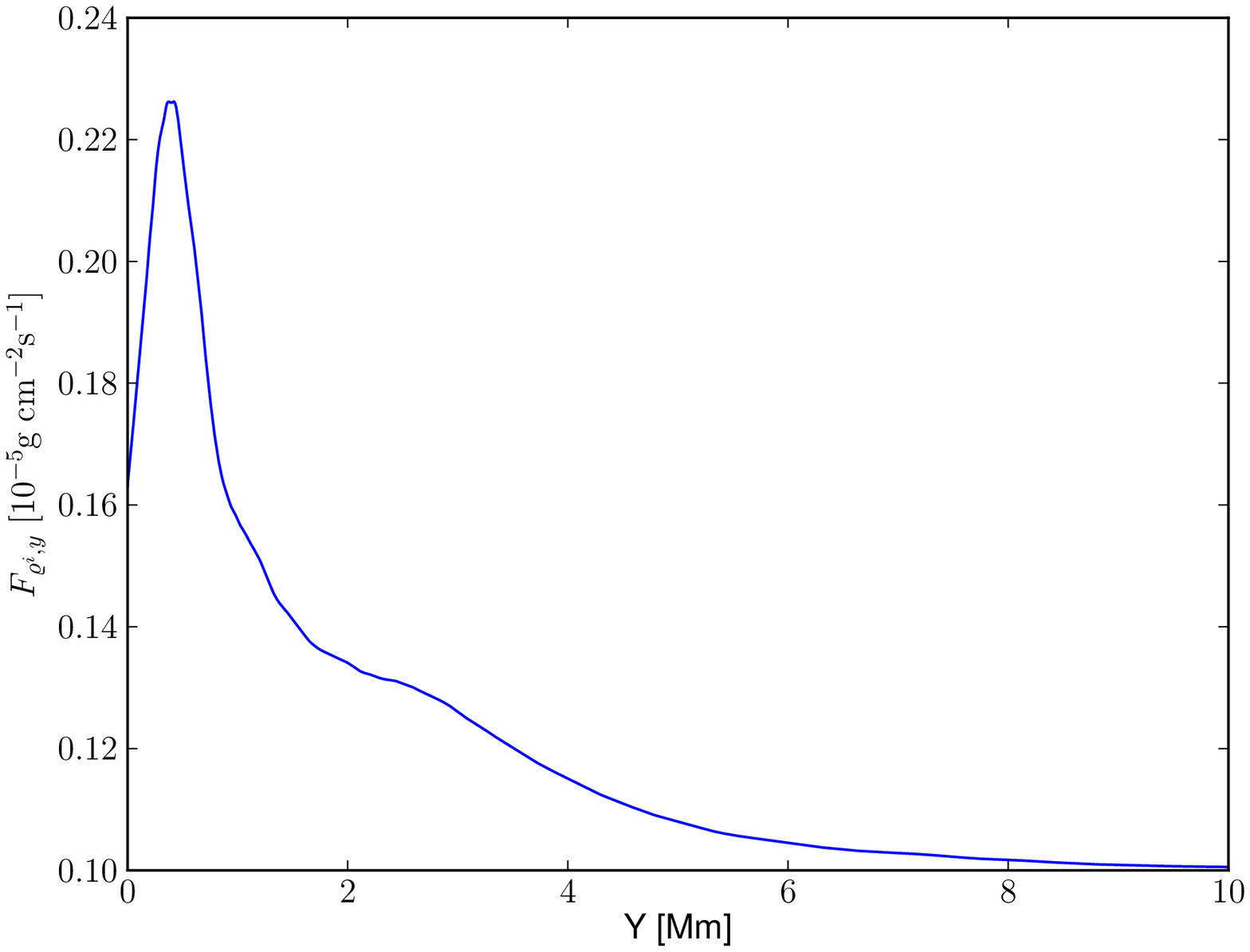}\\ 
\vspace*{1.0cm}\includegraphics[width=\columnwidth]{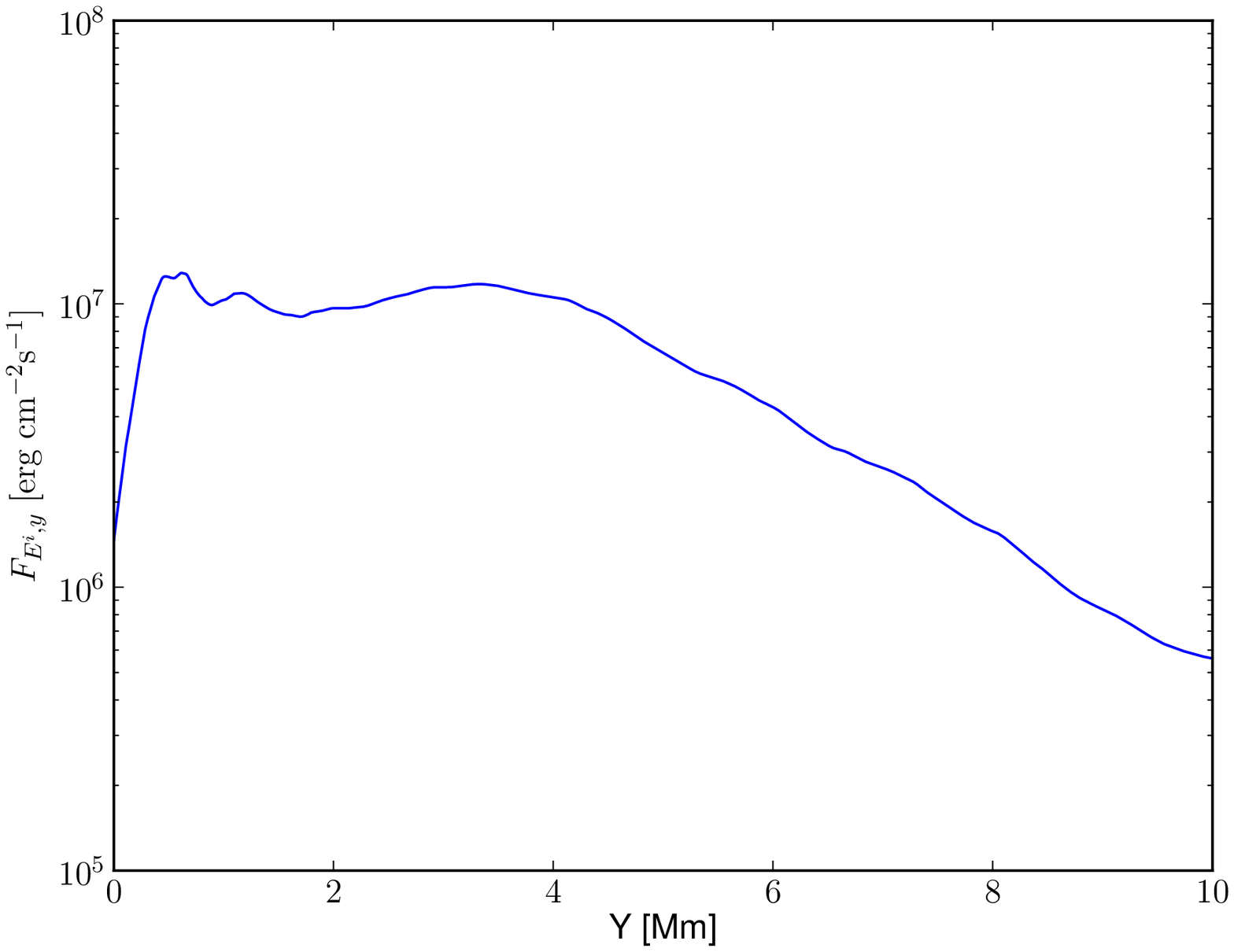}\\
\caption{
Vertical variation of the temporarily and horizontally averaged mass, $<F_{m}(y)>$, (top) 
and energy, $<F_{E}(y)>$, (bottom) fluxes. 
}
\label{fig:energyflux2}
\end{center}
\end{figure}

Figure~\ref{fig:td-plot} displays time-distance plot of vertical component of ion velocity that is averaged over the whole horizontal distance, 
$<V_{{\rm i}\, y}>$. 
The plasma jets emerge from the chromospheric background 
and move into the corona. 
Some of the injected plasma subsides rapidly after reaching its maximum phase \citep[e.g.][]{2017ApJ...849...78K,Srivastava2018}. This entire process is driven by ongoing granulation in the photosphere. 
Analyzing the $<V_{{\rm i}\, y}(y,t)>$ 
we find that the solar corona experiences about $1-3$ minute periods oscillations and $<V_{{\rm i}\, y}>$ reaches a magnitude of $10-20$ km s$^{-1}$ at $y = 8$ Mm and it grows with height. The physical properties of these outflows are aikin to the flow characteristics reported by \cite{Tu2005}. 

Figure~\ref{fig:energyflux} (top) illustrates time-distance plot of the horizontally averaged total vertical ion mass flux, 
$F_{m} (y, t) = <\varrho_{\rm i} V_{{\rm i}\, y} >$, which attains its maximum in the photosphere and lower chromosphere and falls off with height due to rapidly decreasing ion mass density. 
However, even above the transition region 
the estimated 
magnitude of this mass flux lays within the range of $10^{-6}$ and $10^{-5}$ g cm$^{-2}$ s$^{-1}$ and it matches 
the prediction for solar mass losses in the low corona \citep{Withbroe1977}.

Vertical component of ion energy flux transported through the medium can be calculated as 
$F_{\rm E} (x,y,t) = \varrho_{\rm i} \mathbf{V}_{\rm i}^2 V_{\rm iy}/2$.
The time-distance plot of horizontally averaged vertical energy flux, $<F_{\rm E} (y, t)>$, is shown in Fig.~\ref{fig:energyflux} (bottom). 
We see that plasma escaping into the corona above the transition region carries a significant amount of energy (orange and yellow patches). By comparison with time-distance plots of $<V_{{\rm i}\, y}>$ (Fig. \ref{fig:td-plot}) we infer that the energy flux associated with the upflowing plasma is higher than for the descending plasma. 
Note that the obtained values lay within the range of theoretical findings 
for energy losses in the upper chromosphere, transition region and low corona \citep{Withbroe1977}. 

Figure~\ref{fig:energyflux2} shows vertical variation of the temporarily and horizontally averaged mass, $<F_{\rm m}>$, (top) 
and energy, $<F_{\rm E}(y)>$, (bottom) fluxes. 
Note that $<F_{\rm E}(y)>$ grows abruptly within the region of $0<y<0.5$ Mm, where the dense photospheric plasma experiences a push from the below operating granulation. Higher up, that is for $y>0.5$ Mm, $<F_{\rm m} (y)>$ declines with height attaining a value of $10^{-6}$ g cm$^{-2}$ s$^{-1}$ at $y=10$ Mm. 
On the other hand, $<F_{\rm E}(y)>$ remains close to $10^{7}$ erg cm$^{-2}$ s$^{-1}$ in the entire chromosphere and transition region, attaining its local maxima at $y=0.25$ Mm and $y=3.5$ Mm. 
Higher up, it slowly falls off with height, reaching its value of $5 \cdot 10^{5}$ erg cm$^{-2}$ s$^{-1}$ at $y=10$ Mm. 

%
%
\section{Discussion and summary}\label{sec:Summary}
Within the framework of 2-fluid equations for ion-neutral plasma, we performed numerical simulations of 
the origin of the solar wind which results from plasma outflows. These outflows are associated with jets excited by the solar granulation 
which develops in the medium with initially straight magnetic field overlaying 
a hydrostatic equilibrium. 
This configuration well mimics the expanding open magnetic field region in a polar coronal hole. 
Our simulations show that this configuration is later on restructured by granulation 
which operates in the photosphere. 
The whole scenario is associated with the energy and mass leakage into higher atmospheric layers 
in the form of plasma outflows. 
Our results successfully match the expected values of mass and energy losses in the upper chromosphere, 
transition region and low corona \citep{Withbroe1977}. 

It is noteworthy here that \cite{Tu2005} obtained a correlation of the Doppler-velocity and 
radiance maps of spectral lines emitted by various ions (Ne VIII, C IV, Si II) 
with the force-free magnetic field that was extrapolated from the photospheric magnetogram (SOHO/MDI) 
in a polar coronal hole. 
Tu found that Ne VIII ions mostly radiate around the height of 20 Mm above the photosphere, 
where they reveal the outflow speed of about 10 km s$^{-1}$, while C IV ions 
with no average flow speed form essentially around the altitude of 5 Mm. 
Hence, Tu inferred that the plasma outflows start in the coronal funnels 
at altitudes in between 5-20 Mm. 
\cite{Yang2013} 
proposed that magnetic reconnection, 
which took place in the open and closed magnetic field region, 
triggers the plasma outflows observed by \cite{Tu2005}. 
The results of our simulations performed with a novel 2-fluid model of 
a partially-ionized solar atmospheric plasma confirm these observational findings.

Ten years before the plasma outflows were announced by \cite{Tu2005}, 
there was essentially no report on finely-structured jets. 
The exceptions were spicules/macrospicules diversely filling the chromosphere and 
contributing to the mass cycle of the corona \citep[e.g.][]{
Tian2014,WedemeyerBhm2012}. 
In the limit of current observational resolution, it is established 
that the overlaying plasma outflows in the corona must be originated due to 
the contribution from various plasma ejecta. 
Therefore, without emphasizing on a particular type of a jet, 
we simulated the solar granulation which resulted in jets and studied their contribution to 
formation of the solar wind. 

In summary, we investigated formation of plasma outflows between $5$ to $10$~Mm above the photosphere
in the open magnetic field region in a coronal hole as observed by \cite{Tu2005}. 
The outflows in such regions consist of continuous streaming of 
plasma particles from the lower solar atmosphere outward. We point out its linkage to 
the granulation and associated with them the ubiquitous chromospheric jets 
which lead to mass and energy leakage into the inner corona. 
Our model is based on gravitationally stratified and partially-ionized bottom layers of 
the solar atmosphere with adequate temperature and 
magnetic field conditions to mimic the ion-neutral plasma outflows. 
Our studies determine that multiple jets excited by operating in the photosphere granulation are 
able to stimulate continuous plasma outflows in the solar atmosphere which may result in the fast solar wind at higher altitudes in the solar corona. 

\acknowledgments
The authors express their thanks to 
Drs. Ramon Oliver, Roberto Soler and David Martin\'ez-G\'omez 
for their comment on the hydrostatic model of the solar atmosphere.
We would like to thank the reviewer for their time and valuable remarks.
The JOANNA code has been developed by Darek Wójcik. This work was done within the framework of the projects from 
the Polish National Foundation (NCN) Grant nos. 
2017/25/B/ST9/00506 and 
2017/27/N/ST9/01798.
Numerical simulations were performed on 
the LUNAR cluster at Institute of Mathematics of University of M. Curie-Sk{\l}odowska, Lublin, Poland. 
\bibliographystyle{aasjournal}

\end{document}